\documentclass[letterpaper]{JHEP3}
\usepackage{latexsym,amsmath,amsfonts,amssymb,enumerate}
\usepackage[latin1]{inputenc}
\usepackage[pdftex]{graphicx}
\pdfoutput=1
\usepackage{euscript,mathrsfs}
\usepackage{bbm}

\newcommand{\be}{\begin{equation}}
\newcommand{\ee}{\end{equation}}

\def\rref#1{(\ref{#1})}

\DeclareMathOperator{\area}{area}

\DeclareMathOperator{\tr}{tr}

\title{Holographic mutual information is monogamous}

\author{
Patrick Hayden \\
School of Computer Science \& Department of Physics, McGill University, Montreal, Quebec, Canada\\
\email{patrick@cs.mcgill.ca}
}

\author{
Matthew Headrick \\ 
Martin Fisher School of Physics, Brandeis University, Waltham, Massachusetts, USA \\ 
\email{mph@brandeis.edu}
}

\author{
Alexander Maloney \\
Department of Physics, McGill University, Montreal, Quebec, Canada \\
\email{maloney@physics.mcgill.ca}
}

\abstract{We identify a special information-theoretic property of quantum field theories with holographic duals: the mutual informations among arbitrary disjoint spatial regions $A,B,C$ obey the inequality $I(A:B\cup C)\ge I(A:B)+I(A:C)$, provided entanglement entropies are given by the Ryu-Takayanagi formula. Inequalities of this type are known as monogamy relations and are characteristic of measures of quantum entanglement.  This suggests that correlations in holographic theories arise primarily from entanglement rather than classical correlations.  We also show that the Ryu-Takayanagi formula is consistent with all known general inequalities obeyed by the entanglement entropy, including an infinite set recently discovered by Cadney, Linden, and Winter; this constitutes strong evidence in favour of its validity.}

\preprint{BRX-TH-638}

\begin{document}

\section{Introduction}

The holographic principle states that the degrees of freedom of quantum gravity are organized in a way that is consistent with black hole entropy. Thus holography and information theory are intimately related.   
The most precise realization of the holographic principle is the AdS/CFT correspondence, which describes certain theories of quantum gravity in terms of specific quantum field theories living at the boundary of spacetime.  It is therefore natural to ask whether there are purely information-theoretic special properties of quantum field theories with bulk holographic duals.  We will argue that the answer to this question is yes, at least in a suitable large $N$ limit where the bulk theory becomes classical Einstein gravity.

Our starting point is the conjectured formula of Ryu and Takayanagi \cite{Ryu:2006bv,Ryu:2006ef} for entanglement entropies in field theories with holographic duals.  In any quantum field theory, we may choose a region $A$ of space and consider the density matrix $\rho_A$ obtained by tracing over the degrees of freedom outside of this region; its von Neumann entropy $S(A) = -\tr (\rho_A \log \rho_A)$ is called the entanglement entropy of $A$. The density matrix $\rho_A$, and hence $S(A)$, depend on both the choice of region and the state of the field theory.  The Ryu-Takayanagi (RT) conjecture states that, in a field theory with a holographic dual described by classical Einstein gravity, in any state represented by a static spacetime, the entanglement entropy is given by the area of a certain minimal surface:
\be\label{rt}
S(A) = \frac1{4 G_{\rm N}} \min_{M_A}(\area(M_A))\,.
\ee
Here $G_{\rm N}$ is the bulk Newton constant, and the minimum is over surfaces $M_A$ in the bulk that are homologous to $A$ (i.e. $A\cup M_A=\partial a$ for some bulk region $a$). The RT formula satisfies many checks but should be regarded as a conjecture. (See \cite{Nishioka:2009un,Headrick:2010zt} for reviews and discussions.) Classical and quantum corrections to the bulk theory are known to give rise to corresponding corrections to \eqref{rt}; however, the precise form of these corrections is not known in general.

In quantum field theory, the entanglement entropy $S(A)$ contains a short-distance divergence which is proportional to the area of $A$ in the boundary.  In order to remove this divergence we must introduce a regulator, which renders the result scheme-dependent. In the RT formula this is reflected by the fact that  $M_A$ has infinite area, and must be regulated by a choice of cutoff surface in the bulk. In order to obtain a scheme-independent quantity we can construct a linear combination of entanglement entropies for which these short-distance divergences cancel.  For example, given two disjoint, separated regions $A$ and $B$ the quantity
\begin{equation}
I(A:B) := S(A)+S(B)-S(AB)
\end{equation}
(where $AB$ denotes $A\cup B$) is finite and regulator-independent.  This quantity is known as the mutual information.  It measures the total amount of correlation (both classical and quantum) between $A$ and $B$. (See \cite{Groisman2005}, for example, for an operational justification of that statement.) Consistent with this interpretation, the mutual information is always non-negative (\emph{subadditivity of entropy}),
\begin{equation}
I(A:B) \ge 0\,,
\end{equation}
and increases monotonically upon adjoining an extra region $C$ to $B$ (\emph{strong subadditivity of entropy}),
\begin{equation}
I(A:BC) \ge I(A:B)\,.
\end{equation}
Headrick and Takayanagi \cite{Headrick:2007km} showed that the RT formula obeys the strong subadditivity inequality.

In this paper we consider the more complicated \emph{tripartite information} \cite{CasiniHuerta2009} (also called, in the classical information-theory context, the I-measure \cite{Yeung1991} or the interaction information \cite{McGill54})
\begin{align}\label{Idef}
I_3(A:B:C) &:= S(A)+S(B)+S(C) -S(AB)-S(BC)-S(AC)+S(ABC) \notag\\
&=  I(A:B) + I(A:C)-I(A:BC) \,.
\end{align}
The first line makes it clear that $I_3$ is symmetric under permutations of its arguments.\footnote{When the full system is in a pure state, then $I_3$ is in fact symmetric under permutations of $A,B,C,D$, where $D$ is the complement of $ABC$.} It is easy to see that, unlike the mutual information, the area-law divergences in the entanglement entropies cancel in $I_3$ even when the regions share boundaries. The second line of \rref{Idef} shows that $I_3$ can be interpreted as a measure of the ``extensivity" of mutual information.  When $I_3=0$ the mutual information of $A$ with $BC$ is the sum of its mutual informations with $B$ and $C$ individually; the mutual information increases in an extensive manner as we combine $B$ and $C$. In a general quantum system $I_3$ can be either positive, negative, or zero, and a typical quantum field theory will exhibit all three behaviors depending on the choice of $A,B,C$ \cite{CasiniHuerta2009}. $I_3$ is also the combination of entanglement entropies appearing, with a particular configuration of regions, in Kitaev and Preskill's calculation of the topological entanglement entropy of massive theories in three dimensions \cite{KitaevPreskill2006}.

The main result of this paper is that, according to the RT formula, the mutual information is always extensive or superextensive in holographic theories. In other words, for any choice of regions,
\be\label{monogamy}
I_3(A:B:C) \le 0\,.
\ee
The proof is essentially a more elaborate version of the holographic proof of strong subadditivity \cite{Headrick:2007km}. It applies irrespective of the topologies of the bulk (including the possible presence of horizons), the boundary, and the regions $A,B,C$. We further conjecture that \eqref{monogamy} is obeyed by any large-$N$ field theory, whether or not the theory is described by Einstein gravity.

The property \eqref{monogamy} has several interesting implications for holography. First, it provides a novel and strong consistency check on the RT formula.  The entanglement entropies of a quantum system obey various inequalities. Strong subadditivity is one such property; an infinite set of more complicated (and logically independent) inequalities were described in \cite{LindenWinter2005, Cadney2011}. We will show that all of these inequalities follow from \eqref{monogamy}. In fact, it can be shown that all other known identities obeyed by the entanglement entropies are consequences of these combined with some more elementary inequalities.  We conclude that \emph{the RT formula obeys all applicable known general properties of the entanglement entropy}.

Second, assuming that the RT formula is correct, property \eqref{monogamy} is a necessary condition for a field theory to have a classical holographic dual.

Third, the property \eqref{monogamy} may shed some light on the physical nature of correlations in the AdS/CFT correspondence.  In general, the mutual information $I(A:B)$ quantifies both classical correlations and quantum entanglement. AdS/CFT is a strong-weak coupling duality, so one expects that when the bulk theory is classical the boundary theory is highly quantum and the correlations are therefore quantum mechanical in nature. The property \rref{monogamy} can be regarded as a precise version of this statement. To see what this means---and to understand the title of this paper---it is useful to know that inequalities that are structurally of the same form as \eqref{monogamy}, i.e.
\begin{equation}
f(A,B) + f(A,C) \le f(A,BC)\,,
\end{equation}
where $f$ is some measure of entanglement, appear frequently in quantum information theory and quantum cryptography. Such ``monogamy" relations reflect the fact that---unlike classical correlation---entanglement is not a shareable resource: entanglement in the $A$--$B$ system cannot be shared with the $A$--$C$ system. To put it another way, entangled correlations between $A$ and $B$ cannot be shared with a third system $C$ without spoiling the original entanglement. This property is responsible for the security of quantum cryptography. In a general quantum system or quantum field theory, the mutual information does \emph{not} obey the monogamy condition, because it encodes both entanglement and classical correlations. We can thus refer to the property \rref{monogamy} as the statement that in holographic theories mutual information is monogamous. This monogamy suggests that in such theories quantum entanglement dominates over classical correlations.\footnote{Given that the large-$N$ limit is a classical limit, the reader might object to the statement that quantum entanglement could dominate over classical correlations in a large-$N$ theory. It is true that there cannot be entanglement between classical subsystems of a classical system. However, spatial regions of a large-$N$ field theory are not classical subsystems (i.e.\ the division into subsystems does not commute with the classical limit), as evidenced by the fact that they have non-zero entanglement entropies even when the full system is in a pure state.}

In the next section we introduce the tripartite information and describe its behavior in various field theories. A review of the literature shows that, in most cases where explicit calculations are possible, it can take any sign depending on the regions chosen. Section 3 contains our proof that the RT mutual information is monogamous. We also explain why we believe that this result will continue to hold whenever the bulk theory is classical, irrespective of whether it is described by Einstein gravity; in other words, monogamy depends on large-$N$ but not strong coupling. In Section 4 we describe the relationship between monogamy and general inequalities obeyed by the entanglement entropy, showing that monogamy provides evidence for the identification of the RT formula with the entanglement entropy. In Section 5 we give a more physical interpretation of the monogamy property and argue that it is characteristic of quantum as opposed to classical correlations. We conclude in Section 6 with a discussion of open questions.

\section{Tripartite information in quantum field theory}

Before discussing quantum field theories, let us warm up with a 3-qubit system in order to gain intuition for the meaning of the tripartite information \eqref{Idef}. Writing the state vector as $|ABC\rangle$, we consider the following two mixed states (both of which contain purely classical correlations):
\begin{align} \label{3qubitrho}
\rho &= \frac{1}{2} \left( |000\rangle \langle 000 | + | 111 \rangle \langle 111 | \right), \\
\label{3qubittau}
\tau &= \frac{1}{4}  \left( |000\rangle \langle 000 | + | 011 \rangle \langle 011 |+ |101\rangle \langle 101 | + | 110 \rangle \langle 110 | \right).
\end{align}
The state $\rho$ has $I_3(A:B:C) = 1$, because the correlations between $A$ and $B$ are redundant with those between $A$ and $C$. On the other hand, $\tau$ has $I_3(A:B:C)=-1$; $A$ and $B$ are uncorrelated after tracing over $C$, and similarly for $A$ and $C$, while $A$ is perfectly correlated with the joint system $BC$.

It can be shown generally that weak coupling among the subsystems $A,B,C$ leads to non-negative $I_3(A:B:C)$ when the full system $ABC$ is in its ground state. Specifically, for a system with Hamiltonian 
\be
H = H_A \otimes I_{BC} + I_A \otimes H_B \otimes I_C +  I_{AB} \otimes H_C + \lambda H_{ABC},
\ee
a perturbation theory calculation gives
\be
I_3(A:B:C) = -\lambda^2 \log \lambda^2 \; g(H_A,H_B,H_C,H_{ABC}) + \mathcal{O}(\lambda^2)
\ee
where $g \geq 0$~\cite{Mark11}.

The sign of $I_3$ was investigated in a variety of quantum field theories by Casini and Huerta \cite{CasiniHuerta2009}. We will summarize a few of their findings here. As described in the introduction, we always take $A,B,C$ to be disjoint spatial regions. For simplicity, in this section we will restrict ourselves to field theories on Minkowski space in the vacuum.

For a free massless fermion in two dimensions, the entanglement entropy of an arbitrary collection of intervals was computed in \cite{Casini:2004bw,CasiniHuerta2005}, and the result implies $I_3(A:B:C) = 0$ for any $A,B,C$. This is the only case known where the mutual information is always exactly extensive.

For a free massive fermion, Casini and Huerta found that $I_3$ is positive when the sizes and separations of the intervals are small compared to the Compton wavelength, and negative when they are large. Similarly, for a free massive scalar, $I_3$ is positive for small intervals and separations. However, whereas for the fermion it goes to zero in the massless limit (for any fixed set of regions), for the scalar it goes to $+\infty$:
\begin{equation}
I_3 \sim \frac12\log(-\log m)\,.
\end{equation}
The reason is that the long-wavelength modes of the field become zero-modes in this limit; the entanglement entropy of any region then includes a term proportional to the logarithm of the volume of the field space. To put it another way, after tracing over the complement of $ABC$, the constant mode of the field is perfectly classically correlated between the regions, so $\rho_{ABC}$ is essentially of the form \eqref{3qubitrho}.

A similar example is provided by a compactified scalar. For large compactification radius $R$, the mutual information between separated intervals is of the form
\begin{equation}
I(A:B) = \log R + f(A,B)\,,
\end{equation}
where $f(A,B)$ depends on the configuration of $A,B$ but is independent of $R$~\cite{Calabrese:2009ez}. The $\log R$ term is again due to the integration over the zero-momentum mode of the scalar. Hence, if $B$ and $C$ are adjacent, the tripartite information is
\begin{equation}\label{freeboson}
I_3(A:B:C) = \log R+f(A,B)+f(A,C) -f(A,BC)\,,
\end{equation}
which is positive for sufficiently large $R$. 

It is also interesting to consider a general two-dimensional conformal field theory in a different limit. The mutual information between two intervals $A=[u_A,v_A]$, $B=[u_B,v_B]$ is invariant under conformal transformations of the line, so depends only on the cross-ratio
\begin{equation}
x := \frac{(v_A-u_A)(v_B-u_B)}{(u_B-u_A)(v_B-v_A)}\,.
\end{equation}
Calabrese, Cardy, and Tonni \cite{Calabrese:2010he} computed the leading behavior of the mutual information for small $x$ (i.e.\ large separation relative to the sizes of the intervals) in a general CFT:\footnote{It's interesting to note that, parametrically in the separation $r$ between the intervals, the behavior \eqref{ICFT} saturates the quantum Pinsker bound as specialized to connected correlators of normalized operators~\cite{HiaiOT81,Wolf2007}:
\begin{equation}
\left(\frac{\langle\mathcal{O}_A\mathcal{O}_B\rangle - \langle\mathcal{O}_A\rangle\langle\mathcal{O}_B\rangle}{\|\mathcal{O}_A\|\|\mathcal{O}_B\|}\right)^2 \le 2I(A:B)\,.
\end{equation}
We can see this by taking $\mathcal{O}_A,\mathcal{O}_B$ to be the lowest non-unit operator $\hat{\mathcal{O}}$ smeared over the intervals $A,B$ respectively and turned into bounded operators, for example by inserting them into the function $e^{ix}$ \cite{CasiniHuerta2009}:
\begin{equation}
\mathcal{O}_{A,B} = \exp\left(i\int du\,f_{A,B}(u)\hat{\mathcal{O}}(u)\right)\,,
\end{equation}
where $f_{A,B}$ are functions supported on $A,B$ respectively. This will give the largest possible connected correlator for large separations, namely
\begin{equation}
\langle\mathcal{O}_A\mathcal{O}_B\rangle - \langle\mathcal{O}_A\rangle
\langle\mathcal{O}_B\rangle 
 \sim \frac1{r^{2\hat d}}\sim x^{\hat d}\,.
\end{equation}
}
\begin{equation}\label{ICFT}
I(A:B) = \hat m s(\hat d)x^{2\hat d} + \text{(higher order in $x$)}\,,
\end{equation}
where $\hat d$ is the smallest non-zero scaling dimension in the theory, $\hat m$ is its multiplicity, and
\begin{equation}
s(\hat d) = \frac{\pi^{1/2}\Gamma(2\hat d+1)}{4^{2\hat d+1}\Gamma(2\hat d+\frac32)}\,.
\end{equation}
If we now consider a configuration of three intervals $A,B,C$, such that $B,C$ are adjacent and separated from $A$ by a large distance $r$, then we have
\begin{equation}\label{larger}
I_3(A:B:C) = \hat ms(\hat d)\frac{l_A^{2\hat d}}{r^{4\hat d}}\left(l_B^{2\hat d}+l_C^{2\hat d}-(l_B+l_C)^{2\hat d}\right) + (\text{higher order in $1/r$}).
\end{equation}
For this configuration the theory is superextensive, extensive, or subextensive depending on whether $\hat d$ is more than, equal to, or less than $1/2$. This result agrees with the fact that the free fermion, which has $\hat d=1/2$, is always extensive. It also agrees with the fact that the free boson on a large circle, which has $\hat d=1/(2R^2)\ll1$, is subextensive. (Note however that the estimate $I_3\sim\log R$ was obtained in the limit of large $R$ with fixed intervals, whereas \eqref{larger} was obtained in the limit of distant intervals in a fixed theory.)\footnote{We  note that equation \eqref{larger} applies only to CFTs, so is not in conflict with the results of \cite{CasiniHuerta2009} for massive fermions.}

A final interesting example, similar to the two-dimensional massive fermion where $I_3$ is negative for large regions, is provided by three-dimensional massive theories with topological order \cite{KitaevPreskill2006,LevinWen}. The ground state of such a theory is described at long distances by a two-dimensional topological field theory. The local fluctuations of the fields are subject to a constraint that is only apparent at long distances, leading to a finite negative correction to the area law for large regions:
\begin{equation}\label{areacorrection}
S(A) = \alpha\area(\partial A)-\gamma\,\text{comp}(\partial A)\,,
\end{equation}
where $\alpha$ is a UV-divergent quantity, $\gamma$ is finite and nonnegative, and ``comp" means the number of components. By arranging $A,B,C$ as wedges of a disc---so that they and all combinations of them are topologically discs---the area-law divergences cancel and we are left with
\begin{equation}
I_3(A:B:C) = -\gamma\,.
\end{equation}
This quantity is called the \emph{topological entanglement entropy}, and it can be calculated explicitly in terms of basic properties of the theory. The study of examples suggests that \eqref{areacorrection} implies $I_3(A:B:C) \le 0$ for regions with arbitrary topologies. (We do not know a proof of this statement.) Hence such theories apparently obey the monogamy property, at least for regions large enough that \eqref{areacorrection} applies. If this is correct, it would be very interesting to understand whether there is any connection to the monogamy property of holographic theories that we study in this paper.

Using the Ryu-Takayanagi formula, Pakman and Parnachev computed the topological entanglement entropy in a general confining holographic theory, showing that it always vanishes \cite{Pakman:2008ui}. The reason is that the dual gravitational theory, being classical and local, is incapable of describing topological order. However, it should be kept in mind that the RT formula captures only the order $G_{\rm N}^{-1}\sim N^2$ part of the entanglement entropy, so Pakman and Parnachev's result does not preclude the possibility of an order--1 topological entanglement entropy.

\section{Holographic mutual information is monogamous}

\subsection{Proof}

We now describe the main result of this paper, which is that the Ryu-Takayanagi entropies obey the inequality
\begin{equation}\label{superextensivity}
I_3(A:B:C) \le 0
\end{equation}
for any regions $A,B,C$ in the boundary field theory. The proof is similar to the proof of strong subadditivity by Headrick and Takayanagi \cite{Headrick:2007km}. The RT formula applies to field theories that possess holographic duals described by classical Einstein gravity. In the next subsection we will consider the effect of higher-curvature corrections to the bulk gravitational action, as well as more general large-$N$ field theories. In subsection 6.2 we will discuss finite-$N$ corrections, i.e.\ quantum effects in the bulk.

We begin by setting up some notation and recalling the precise form of the RT formula. The RT formula applies to states that are represented in the dual by static classical geometries, and gives the entanglement entropy of a spatial region of the field theory in terms of the area of a minimal surface lying on a constant-time slice of the bulk geometry. The time direction plays no role so we will suppress it. We will denote the constant-time slice of the boundary by $X$; this is the space on which the field theory lives. The constant-time slice of the bulk is $Y$ (so $X\subset\partial Y$; however, $Y$ may also have ``internal" boundaries, such as horizons or walls). The boundary of any bulk region $r$ can be decomposed into a part lying in $X$ and a part lying in $Y$. (For this purpose we consider the part of $\partial r$ lying along an internal boundary of $Y$ to be ``lying in $Y$".) We can therefore define the two boundary operators $\partial_X, \partial_Y$ by
\begin{equation}
\partial r = \partial_Xr\cup\partial_Yr\,,\qquad
\partial_Xr := \partial r\cap X\,,\qquad
\partial_Yr := \partial r\cap Y\,.
\end{equation}
The RT formula is
\begin{equation}\label{RT}
S(A) = \frac1{4G_{\rm N}}\mathop{\min_{\tilde a\subset Y:}}_ {\partial_X\tilde a=A}\left(\area(\partial_Y\tilde a)\right).
\end{equation}
The area is computed with respect to the spatial part of the Einstein-frame metric. We will refer to the minimizing region for $A$ as $a$, the one for $AB$ as $ab$, etc. Note  that, whereas $AB:=A\cup B$, in general $ab\neq a\cup b$.

Our strategy for proving \eqref{superextensivity} will be as follows. We will divide the three surfaces $\partial_Yab$, $\partial_Ybc$, and $\partial_Yac$  into four pieces each. We will then reassemble these twelve surfaces into four surfaces which are the boundaries of certain regions $\tilde a$, $\tilde b$, $\tilde c$, and $\widetilde{abc}$ respectively, with $\partial_X\tilde a=A$ etc. Hence
\begin{align}\label{strategy}
S(AB)+S(BC)+S(AC) &= \frac1{4G_{\rm N}}\left(\area(\partial_Yab)+\area(\partial_Ybc)+\area(\partial_Yac)\right) \nonumber \\
&= \frac1{4G_{\rm N}}\left(\area(\partial_Y\tilde a)+\area(\partial_Y\tilde b)+\area(\partial_Y\tilde c)+\area(\partial_Y\widetilde{abc})\right) \nonumber \\
&\ge S(A)+S(B)+S(C)+S(ABC)\,,
\end{align}
where in the last line we used the fact that actual entropy minimizes the area for each region. Figure \ref{fig:recolor} gives graphical depiction of the strategy.

\FIGURE{
\includegraphics[width=6in]{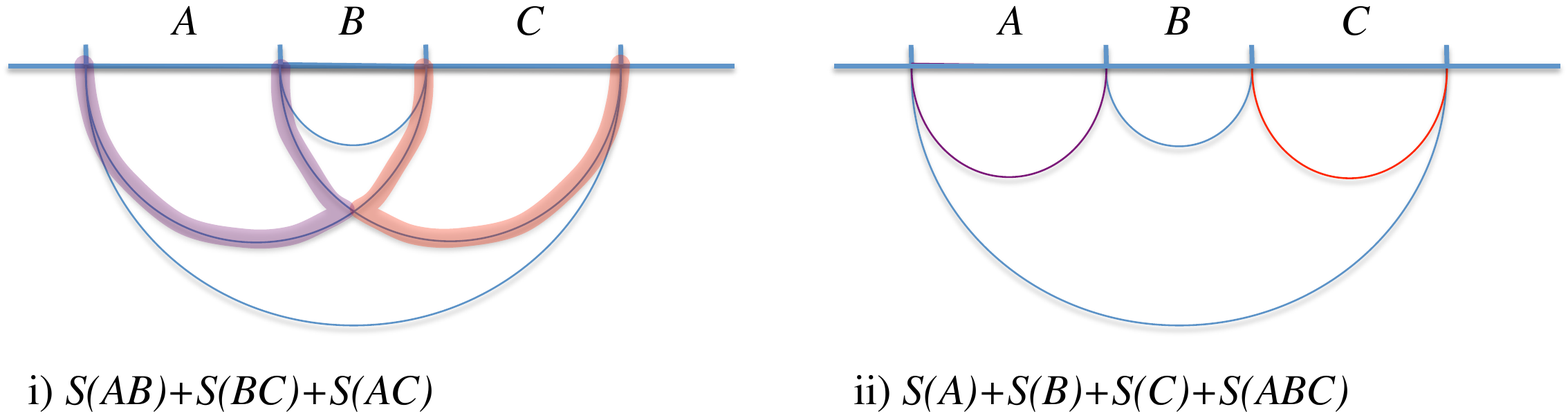}
\caption{Illustration of the proof in AdS${}_3$/CFT${}_2$. (i) The three minimal surfaces corresponding to $S(AB)$, $S(BC)$, and $S(AC)$, with the latter consisting of two connected components in the bulk. (ii) The four minimal surfaces corresponding to $S(A)$, $S(B)$, $S(C)$, and $S(ABC)$. In this example, the surfaces for $S(AB)$ and $S(BC)$ each get decomposed into a purple part and a red part. The purple forms a surface terminating on the boundary of $A$ and the red a surface terminating on the boundary of $C$. Since these new surfaces need not be minimal, their areas form upper bounds on $S(A)$ and $S(C)$, respectively, by the Ryu-Takayanagi formula. The reader is cautioned, however, that these simple cartoons can be misleading. As we will see in section 6.1,  higher-order versions of the monogamy inequality fail even though naive graphical arguments suggest they should hold.} \label{fig:recolor}
}

We will decompose the surface $\partial_Yab$ into four parts: the part in both $ac$ and $bc$; in $ac$ but not $bc$; in $bc$ but not $ac$; and in neither $ac$ nor $bc$:
\begin{equation}\label{abbound}
\partial_Yab =
\left(\partial_Yab\cap bc\cap ac\right)\cup
\left(\partial_Yab\cap bc\setminus ac\right)\cup
\left(\partial_Yab\cap ac\setminus bc\right)\cup
\left(\partial_Yab\setminus bc\setminus ac\right)\,.
\end{equation}
Similarly for $bc$ and $ac$:
\begin{equation}\label{bcbound}
\partial_Ybc =
\left(\partial_Ybc\cap ab\cap ac\right)\cup
\left(\partial_Ybc\cap ab\setminus ac\right)\cup
\left(\partial_Ybc\cap ac\setminus ab\right)\cup
\left(\partial_Ybc\setminus ab\setminus ac\right)\,,
\end{equation}
\begin{equation}\label{acbound}
\partial_Yac =
\left(\partial_Yac\cap bc\cap ab\right)\cup
\left(\partial_Yac\cap bc\setminus ab\right)\cup
\left(\partial_Yac\cap ab\setminus bc\right)\cup
\left(\partial_Yac\setminus bc\setminus ab\right)\,.
\end{equation}
We now define the following regions:
\begin{align}
\tilde a &= ab\cap ac\setminus bc \nonumber \\
\tilde b &= ab\cap bc\setminus ac \nonumber \\
\tilde c &= ac\cap bc\setminus ab \nonumber \\
\widetilde{abc} &= ab\cup bc\cup ac\,.
\end{align}
As required, these regions are anchored on $A$, $B$, $C$, and $ABC$ respectively:
\begin{equation}
\partial_X\tilde a = A\,,\qquad\quad
\partial_X\tilde b = B\,,\qquad\quad
\partial_X\tilde c = C\,,\qquad\quad
\partial_X\widetilde{abc} = ABC\,.
\end{equation}
In the bulk, we have
\begin{align}\label{rearranged}
\partial_Y\tilde a &= \left(\partial_Yab\cap ac\setminus bc\right)\cup \left(\partial_Yac\cap ab\setminus bc\right)
\cup\left(\partial_Ybc\cap ab\cap ac\right) \nonumber \\
\partial_Y\tilde b &= \left(\partial_Yab\cap bc\setminus ac\right)\cup \left(\partial_Ybc\cap ab\setminus ac\right)
\cup\left(\partial_Yac\cap ab\cap bc\right) \nonumber \\
\partial_Y\tilde c &=  \left(\partial_Yac\cap bc\setminus ab\right)
\cup\left(\partial_Ybc\cap ac\setminus ab\right)\cup\left(\partial_Yab\cap ac\cap bc\right) \nonumber \\
\partial_Y\widetilde{abc} &= \left(\partial_Yab\setminus bc\setminus ac\right)\cup \left(\partial_Yac\setminus ab\setminus bc\right)\cup\left(\partial_Ybc\setminus ab\setminus ac\right).
\end{align}
By inspection, each term on the right-hand side of \eqref{rearranged} equals precisely one term on the right-hand side of \eqref{abbound}, \eqref{bcbound}, \eqref{acbound}.
This establishes the second equality of \eqref{strategy}, and completes the proof.

\subsection{Higher-curvature corrections and general large-$N$ theories}

We now consider the effect of higher-curvature terms in the bulk gravitational action, such as $\alpha'$ corrections in string theory. (In this subsection we stay in the classical---i.e.\ large-$N$---limit in the bulk; quantum effects are discussed in subsection 6.2.) It has been conjectured \cite{Fursaev:2006ih,Headrick:2010zt} that, in the presence of such corrections, the entanglement entropy is still given by a minimization over surfaces, but with the area functional itself corrected by higher-derivative terms:
\begin{align}\label{higherderivative}
S(A) &= \frac1{4G_{\rm N}}\mathop{\min_{\tilde a\subset Y:}}_ {\partial_X\tilde a=A}\left(F(\partial_Y\tilde a)\right), \\ \label{small}
\qquad F &= \area \; + \; \text{higher derivative terms}.
\end{align}
Such terms could include, for example, an Einstein-Hilbert term for the induced metric. A general formula for the functional $F$ has not been proposed, but formulas applying for certain types of corrections to the gravitational action have been proposed and tested in \cite{deBoer:2011wk,Hung:2011xb}. In fact, we can go further and conjecture that, even when there is no Einstein-Hilbert term--- such as in topological or conformal theories of gravity --- the entanglement entropy is given by \eqref{higherderivative}, but where $F$ is not necessarily of the form \eqref{small}. Thus, the bulk need not even be geometrical, in the sense of being equipped with a Riemannian metric; it just needs to be a topological space on which one can define regions and their boundaries.

One general condition on the functional $F$ appearing in \eqref{higherderivative} can be deduced from the strong subadditivity property. The proof that the RT formula satisfies strong subadditivity goes through for the formula \eqref{higherderivative}, if and only if $F$ is extensive (i.e.\ for disjoint surfaces $s_1,s_2$, $F(s_1\cup s_2) = F(s_1)+F(s_2)$) \cite{Headrick:2007km}. The same holds for the proof of monogamy given in the previous subsection. We can therefore conclude that, in the language of the boundary field theory, the monogamy property requires large $N$ but not strong coupling.

We can go a step further, and conjecture more speculatively that the monogamy property holds in any large-$N$ field theory.  More precisely, we expect that the leading (order $N^2$) part of the entanglement entropy is monogamous in any field theory that becomes classical in its large-$N$ limit (in the sense of \cite{Yaffe:1981vf}).  The main evidence for this statement, aside from an extrapolation of the previous line of argument, comes from the results of \cite{Headrick:2010zt}. There it was argued that a general large-$N$ two-dimensional CFT (such as the symmetric-product orbifold theory $\mathcal{C}^N/S_N$, in the large-$N$ limit, where $\mathcal{C}$ is an arbitrary compact unitary CFT) has the same entanglement entropies as holographic ones, for arbitrary spatial regions (and even the same entanglement R\'enyi entropies). All of these theories therefore satisfy the monogamy property.

All of this suggests that entanglement entropies in large-$N$ theories may have a simple, universal structure, and therefore their study may be very fruitful.\footnote{Entanglement entropies in the $O(N)$ model at large $N$ were studied in the paper \cite{Sachdev}.}

\section{Ryu-Takayanagi entropy as entanglement entropy}

We now describe the relationship between the monogamy inequality $I_3\le0$ and other inequalities obeyed by entanglement entropies.
 
For a general quantum system, the entanglement entropies will always obey certain inequalities.
For example,
\begin{enumerate}[\quad(a)]
\item \emph{Subadditivity}: $S(A)+S(B)\ge S(AB)$
\item \emph{Araki-Lieb}: $S(AB)\ge |S(A)-S(B)|$
\item \emph{Strong subadditivity 1}: $S(AB)+S(BC)\ge S(ABC)+S(B)$
\item \emph{Strong subadditivity 2}: $S(AB)+S(BC)\ge S(A)+S(C)$.
\end{enumerate}

Headrick and Takayanagi proved that all four of these inequalities are satisfied by the RT formula \cite{Headrick:2007km}.\footnote{In the context of formal quantum information theory, inequalities (a) and (b) are equivalent, as are (c) and (d), but when proved using the RT formula, they are logically independent. The reason is that (a) and (b) (or (c) and (d)) can be transformed into each other by considering a purification of the mixed state in question. That trick is not available in the holographic context for a thermal state at finite temperature because the purified state need not have a simple geometric interpretation. } This constitutes a consistency check of the RT formula, rather than a characterization of any special property of holographic theories.  
On the other hand, the monogamy property
\begin{enumerate}[\quad($\star$)]
\item \emph{Monogamy}: $S(AB)+S(AC)+S(BC)\ge S(ABC)+S(A)+S(B)+S(C)$,
\end{enumerate}
is not obeyed by a general quantum system.  This property is logically independent of the inequalities (a)--(d) and represents a novel constraint on any theory that is conjectured to possess a holographic dual.

In addition to the well-known inequalities (a)--(d), Linden and Winter~\cite{LindenWinter2005}, and subsequently Cadney, Linden, and Winter~\cite{Cadney2011}, have shown that entanglement entropies obey additional constrained inequalities. These inequalities are independent of (a)--(d) in the sense that there exist functions which satisfy (a)--(d) but which violate the constrained inequalities. We will demonstrate, however, that (a)--(d) supplemented with monogamy ($\star$)  imply the full suite of constrained inequalities, providing an additional check on the consistency of the RT formula.

The constrained inequalities are most easily expressed in terms of the conditional mutual information
\be
I(A:C|B) := S(AB) + S(BC) - S(ABC) - S(B) = I(A:BC)-I(A:B)\,.
\ee
In terms of this new quantity, strong subadditivity (c) is simply $I(A:C|B) \ge 0$ while monogamy ($\star$) can be written as $I(A:C|B) \ge I(A:C)$.

\begin{enumerate}[\quad(a)]
\setcounter{enumi}{4}
\item \emph{Linden-Winter~\cite{LindenWinter2005}}:
If 
\begin{equation}\label{condition}
I(A:C|B) = I(A:B|C) = I(B:C|D) = 0
\end{equation}
then $I(C:D) \geq I(C:AB)$.
\end{enumerate}

It is easy to show that the RT formula obeys the Linden-Winter condition.  Consider a set of four regions obeying the constraint \rref{condition}.  Monogamy ($\star$) implies that $I(B:C|D) \ge I(B:C)$ but \rref{condition} requires that $I(B:C|D) = 0$, while property (a) specifies that $I(B:C) \geq 0$. With bounds above and below, we are left with $I(B:C) = 0$. 
Meanwhile, a trivial identity which follows immediately from the definitions states that $I(C:AB)$ can be written as
\be
I(C:AB) = I(B:C) + I(A:C|B) 
\ee and using the constraint $I(A:C|B) = 0$, we see that $I(C:AB)=0$.  Thus the inequality $I(C:D) \geq I(A:BC)$ is satisfied, albeit in a rather trivial way.  

\begin{enumerate}[\quad(a)]
\setcounter{enumi}{5}
\item \emph{Cadney-Linden-Winter~\cite{Cadney2011}}:
If  $I(A:C|B) = I(B:C|A) = 0$, then
\be
S(X_1\ldots X_n) + (n-1) I(AB:C) \leq \sum_{i=1}^n S(X_i) + \sum_{i=1}^n I(A:B|X_i)
\ee
for any quantum state and disjoint subsystems $\{A, B, C, X_1, \ldots, X_n \}$.
\end{enumerate}

The argument that the RT formula implies this inequality is similar. By the same reasoning as before, the constraint implies that $I(A:C) = 0$. But $I(AB:C) = I(A:C) + I(B:C|A)$, which is zero by another application of the constraint. Since $I(A:B|X_i) \geq 0$ by strong subaddivity (c) and $S(X_1 \ldots X_n) \leq S(X_i)$ by repeated applications of subadditivity (a), the inequality follows.

The article~\cite{Cadney2011} contains three variations on this family of inequalities, which are proved by purifying the quantum state, applying (f) and relabelling in different ways. The RT formula implies these additional inequalities by very similar arguments to the ones we've already given. We omit the details, the only significant modification being that the Araki-Lieb inequality (b) occasionally substitutes for subadditivity (a).

Inequalities (a)--(d) along with (e),(f) together imply all known general inequalities obeyed by the entanglement entropy \cite{LindenWinter2005, Cadney2011} so  we conclude that the RT entropy is consistent with all known inequalities obeyed by the entanglement entropy.

\section{Holography and the structure of correlations}

In this section we consider implications of the monogamy inequality for the structure of 
correlations in holographic theories, first by showing that quantum Markov chains are 
forbidden and then by considering other situations in which monogamy arises in quantum information theory.

\subsection{Absence of quantum Markov chains}

The random variables $X$, $Y$, and $Z$ form a Markov chain $X - Y - Z$ if $p(x,y,z)$ has the form $p(x|y)p(z|y)p(y)$, that is, $X$ and $Z$ are conditionally independent given $Y$. The interpretation is that any correlations between $X$ and $Z$ are mediated by $Y$. The condition can equivalently be written $p(x,y,z)=p(x,y)p(z|y)$, saying that there is a stochastic map taking $Y$ to $Z$ in such a way as to correctly reproduce all correlations between $X$ and $Z$.
A simple way to test if the triple forms a Markov chain is to evaluate some entropies: $I(X:Z|Y) = 0$ if and only if $X - Y - Z$ forms a Markov chain.

For quantum states, the condition $I(A:C|B) =  0$ identifies an analogous set of states known as \emph{quantum Markov chains}. A quantum Markov chain of the systems $A$, $B$, and $C$ is a quantum state $\rho_{ABC}$ such that there is an open system evolution map (completely positive, trace-preserving linear map) $\Gamma$ from the density operators on $B$ to those of $BC$ such that $\rho_{ABC} = (I_A \otimes \Gamma)\rho_{AB}$~\cite{Petz86}. For such a state, the correlations between $A$ and $C$ are mediated by $B$ via the map $\Gamma$, which is essentially just a noncommutative generalization of $p(z|y)$ from the random variable case.

The monogamy inequality $I(A:C|B) \geq I(A:C)$ implies that when $I(A:C) \neq 0$, one must have $I(A:C|B) > 0$, so the state $\rho_{ABC}$ cannot be a quantum Markov chain. It follows that quantum Markov chains are prohibited by holography when $A$, $B$ and $C$ represent spatial regions.

In the finite-dimensional case, all quantum Markov chain states have the form~\cite{HaydenJPW04}
\begin{equation} \label{eqn:qmarkov}
\rho_{ABC} = \sum_j p_j \rho_{AB_j^L} \otimes \rho_{B_j^RC},
\end{equation}
where the Hilbert space $B$ has an orthogonal direct sum decomposition $B \cong \oplus_j (B_j^L \otimes B_j^R)$ and $p$ is a probability vector. (The validity of a similar decomposition in the infinite-dimensional setting is under investigation~\cite{Shirokov11}.) 
The state $\rho_{ABC}$ in \rref{3qubitrho} is a particularly simple example. An immediate
consequence of \rref{eqn:qmarkov} is that no quantum Markov chain contains any entanglement between $A$ and $C$. 

Relativistic vacuum states, however, are known to be highly entangled. In fact, under very general assumptions, the state of any pair of disjoint spatial regions will even violate a Bell inequality~\cite{SummersWerner87}. The vacuum therefore doesn't admit spatial quantum Markov chains. 
Monogamy, which prohibits spatial quantum Markov chains in holographic field theories whenever $I(A:C) > 0$, implies that the conclusion continues to hold at arbitrarily high temperature.

\subsection{Monogamy and quantum cryptography} \label{sec:crypto}

The monogamous nature of truly quantum-mechanical correlations is actually one of their most useful properties from the point of view of information theory and cryptography. Given a pure, entangled quantum state $|\varphi\rangle_{AB}$ of the systems $A$ and $B$, the only states of $A$, $B$ and $C$ consistent with $\varphi$ all have the form
\begin{equation}
\rho_{ABC} = |\varphi\rangle\langle\varphi|_{AB} \otimes \sigma_C,
\end{equation}
with absolutely no correlation between $AB$ and $C$. Having good entanglement between $A$ and $B$ makes it impossible for $A$ and $C$ to share entanglement or even weaker forms of correlation, hence the term monogamy. In quantum cryptography, this property is exploited to ensure the security of correlations established between $A$ and $B$, with $C$ playing the role of an eavesdropper. If an experimental procedure can establish that the state shared by $A$ and $B$ is close to a pure entangled state, then the eavesdropper $C$ cannot learn anything about the correlations between $A$ and $B$, which can then be used as a cryptographic secret key.

The monogamy property is manifested mathematically through inequalities relating various measures of correlation. Unlike the simple situation of a pure state partitioned into two halves, for which the entanglement entropy is in a rigorous sense the unique asymptotic measure of entanglement, for mixed states there are many measures that in general do not coincide~\cite{HHH2000}. That is not a defect of the theory so much as a reflection of the fact that mixedness introduces unavoidable irreversibility into the theory, leading to gaps between quantities that agree in the simple pure state setting.

The entanglement of formation~\cite{BennettDSW96} is defined as
\begin{equation}
E_f(A:B)_\rho = \inf \left\{ \sum_i p_i S(A)_{\psi_i} 
	: \rho = \sum_i p_i |\psi_i\rangle\langle\psi_i|_{AB} \right\}.
\end{equation}
The subscripts $\rho$ and $\psi_i$ identify the state with respect to which the given function should be evaluated, while the infimum is over all ways of decomposing $\rho$ into a convex combination of pure states. Unlike for probability distributions, this decomposition is far from unique. $E_f$ is related to the amount of entanglement required to produce $\rho$: given a decomposition $\rho = \sum_i p_i |\psi_i\rangle\langle\psi_i|_{AB}$, one could prepare the state $|\psi_i\rangle_{AB}$ with probability $p_i$, leading to an expected entanglement investment of $\sum_i p_i S(A)_{\psi_i}$. Minimizing over decompositions leads to more efficient preparation procedures. (For an investigation of the entanglement of formation in quantum field theories, see~\cite{Narnhofer02}. There are subtleties in the definition and regularization of the entanglement measures discussed here that we are ignoring for the sake of clarity in this overview.)

As suggested by the example at the beginning of this section, strong entanglement between $A$ and $B$ should limit all forms of correlations between $A$ and $C$. The amount of correlation that can be extracted at $C$ can be quantified by introducing a measurement procedure for $C$ and evaluating the mutual information between the measurement outcomes and the system $A$. Any measurement is characterized by a positive operator-valued measure: a set of operators $\{ M_x \}$ with $M_x$ positive semidefinite and $\sum_x M_x = I$. (We restrict ourselves for simplicity to measurements with a finite number of outcomes.) The probability of outcome $x$ is $p_x = \tr \rho_C M_x$ and the state on $A$ conditioned on the outcome $x$ occurring is $\rho_A^{(x)} = \tr_C [ (I_A \otimes M_x) \rho_{AC} ] / p_x$. Defining the post-measurement state $\tau$, we have
\begin{equation}
\tau_{AX} = \sum_x p_x \rho_A^{(x)} \otimes |x\rangle\langle x|_X
\end{equation}
and can calculate
\begin{equation}
I(A:X)_\tau = S(\rho_A) - \sum_x p_x S( \rho_A^{(x)} ).
\end{equation}
Maximizing this mutual information over all measurement procedures leads to a measure $I^\leftarrow(A:C)_\rho$ of the correlation between $A$ and $C$ which is constrained by monogamy, the larger $E_f(A:B)$ is, the smaller $I^\leftarrow(A:C)$ must be:
\begin{equation} \label{eqn:Ef_monogamy}
E_f(A:B)_\rho + I^\leftarrow(A:C)_\rho \leq S(A)_\rho,
\end{equation}
with equality if the state $\rho_{ABC}$ is pure~\cite{KoashiWinter03}. 

At the other end of the spectrum of entanglement measures, there is the distillable entanglement, which summarizes how useful the state $\rho_{AB}$ is, instead of how costly it is to produce as was the case with $E_f$. Specifically, the distillable entanglement quantifies the rate at which standard entangled states $|00\rangle_{AB} + |11\rangle_{AB}$ that can be extracted from the state $\rho_{AB}^{\otimes n}$ in the limit of many copies~\cite{BennettDSW96,DevetakWinter05}. In the case of distillation procedures involving arbitrary local manipulations of the $A$ and $B$ systems, supplemented by the communication of measurement outcomes from $B$ to $A$, the resulting optimal rate $E_D^\leftarrow$ satisfies~\cite{KoashiWinter03}
\begin{equation} \label{eqn:Ed_monogamy}
E_D^\leftarrow(A:B) + E_D^\leftarrow(A:C) \leq E_D^\leftarrow(A:BC).
\end{equation}

Inequalities (\ref{eqn:Ef_monogamy}) and (\ref{eqn:Ed_monogamy}) are structurally very similar to the mutual information mo\-nog\-a\-my inequality
\begin{equation} \label{eqn:I_monogamy}
I(A:B)_\rho + I(A:C)_\rho \leq I(A:BC)_\rho.
\end{equation}
Indeed, if $\rho_{ABC}$ is pure, then $S(A)_\rho = I(A:BC)_\rho / 2$, making the similarity even more pronounced for Eq.~(\ref{eqn:Ef_monogamy}). There is a crucial difference, however. The monogamy relations for $E_f$ and $E_D^\leftarrow$ hold for any quantum mechanical state whereas counterexamples to monogamy of mutual information abound. The monogamy of mutual information in holographic field theories therefore implies that the correlations in the theory are very special. It is tempting to speculate that they are special in the sense of being highly quantum mechanical or even that the mutual information is primarily assessing entanglement.

One way to formulate that question rigorously involves a third (and final!) entanglement measure, known as the squashed entanglement~\cite{ChristandlW04}. Unlike the entanglements of formation and distillation, squashed entanglement is popular because of its convenient abstract properties, not because of a compelling operational interpretation. The definition is
\begin{equation} \label{eqn:Esq_defn}
E_{sq}(A:B)_\rho = \frac{1}{2} \inf_{\rho'_{ABE}} I(A:B|E)_{\rho'},
\end{equation}
where the infimum is over all states $\rho'$ on $ABE$ that agree with $\rho$ on $AB$. $E$, in turn, can be an arbitrary quantum system. Any pure $\rho_{AB}$ only has extensions of the form $\rho'_{ABE} = \rho_{AB} \otimes \sigma_E$, for which $I(A:B|E)_{\rho'} = I(A:B)_{\rho'} = 2 S(A)_\rho$. The squashed entanglement therefore reduces to the entanglement entropy for pure states. On the other hand, unentangled mixed states, known as separable states, have the form
\begin{equation}
\rho_{AB} = \sum_i p_i \rho_A^{(i)} \otimes \rho_B^{(i)}.
\end{equation}
Such states all have extensions of the form
\begin{equation}
\rho'_{ABE} = \sum_i p_i \rho_A^{(i)} \otimes \rho_B^{(i)} \otimes |i\rangle\langle i|_E,
\end{equation}
for which $I(A:B|E) = 0$ because, conditioned on $i$ (the contents of $E$), the $AB$ system is in a product state. Thus, $E_{sq}$ is zero for unentangled states. Moreover, it was recently shown using an ingenious argument that $E_{sq}$ is always strictly positive for entangled states~\cite{BrandaoCY10}.

Monogamy of mutual information, however, is equivalent to the inequality $I(A:B|C)_\rho \geq I(A:B)$ for all disjoint spatial regions $A$, $B$ and $C$. Therefore, $\inf_C I(A:B|C)_\rho = I(A:B)_\rho$. If it were sufficient to restrict the infimum of Eq.~(\ref{eqn:Esq_defn}) to an infimum over extensions corresponding to spatial regions, we would be able to conclude that $E_{sq}(A:B)_\rho = I(A:B)_\rho /2$. That is, \emph{all} the correlation in the holographic quantum field theory could be attributed to entanglement.

Unfortunately, monogamy of mutual information is not sufficient to attribute the bipartite correlations to entanglement alone.
As we have seen, certain classical probability distributions yield monogamous mutual informations without the need for any entanglement. The state $\tau_{ABC}$ of Eq.~(\ref{3qubittau}) provides a simple example. At zero temperature, however, the state of the quantum field theory must be pure whereas $\tau_{ABC}$ is mixed. It is therefore natural to ask whether a  pure quantum state with parts correlated in an essentially classical way is consistent with the monogamy of mutual information. Generalizing the simple example, one could consider any states of the form
\begin{equation}
\tau_{ABC} = \sum_{xyz} p(x,y,z) | x \rangle\langle x |_A \otimes | y \rangle\langle y |_B
		\otimes | z \rangle\langle z |_C.
\end{equation}
The simplest way to write down a pure state consistent with such a $\tau_{ABC}$ is
\begin{equation}
|\tau\rangle_{AA'BB'CC'} = \sum_{xyz} \sqrt{p(x,y,z)} |x\rangle_A |x\rangle_{A'} |y\rangle_B 
			|y\rangle_{B'} |z\rangle_C |z\rangle_{C'}.
\end{equation}
Monogamy of the mutual information for $\tau_{ABC}$, however, need not extend to the pure state $|\tau\rangle_{AA'BB'CC'}$. In particular, a short calculation exploiting the purity of the overall state gives
\begin{eqnarray*}
I_3(A:A':BC)_\tau 
&=& S(A)_\tau - 2 \left[ S(ABC)_\tau - S(BC)_\tau \right].
\end{eqnarray*}
For a choice of state such as that in Eq.~(\ref{3qubittau}), $S(ABC)_\tau = S(BC)_\tau$ since the $A$ bit $x$ is completely determined by the values of the bits $y$ and $z$ of $B$ and $C$, leading to a violation of monogamy. 

Of course, this is not the only way to write down a purification of $\tau_{ABC}$. Other less symmetric choices can be used to recover monogamy of the mutual information, at least in the case of the state of Eq.~(\ref{3qubittau}). In that case, $z = x+y \mod 2$ so the $C'$ system is not necessary to construct the purifying state. The resulting purification
\begin{equation}
\frac{1}{2} \sum_{xy}  |x\rangle_A |x\rangle_{A'} |y\rangle_B 
			|y\rangle_{B'} | z \rangle_C 
\end{equation}
\emph{does} satisfy monogamy of the mutual information for all triples of subsystems.

Thus, while the monogamy of mutual information is consistent with the conclusion that the bipartite correlations in a holographic field theory are dominated by entanglement, it does not imply the conclusion. We nonetheless speculate that the squashed entanglement between any spatial regions $A$, $B$, and $C$ is bounded below by a constant (independent of the regions) times the mutual information.

\section{Open questions}

We have established that the mutual information is monogamous in holographic field theories, assuming the Ryu-Takayanagi formula, and we have made two conjectures regarding the range of validity and interpretation of our result. First, we suspect that the monogamy property should hold to leading order in $N^2$ in any large-$N$ field theory, not just in theories with holographic duals. Second, we believe that the monogamy property of the mutual information in these theories indicates that the mutual information is detecting quantum mechanical correlations in the form of entanglement. One possible way to quantify that statement can be found in section \ref{sec:crypto}, but there are many other possibilities. We challenge the reader to establish quantitatively
that correlations in holographic field theories are dominated by entanglement.

We conclude with a few additional open questions. 

\subsection{Additional inequalities} \label{sec:additional_ineq}

It is natural to ask whether the RT formula obeys other inequalities.  The known properties apply to one, two and three regions, respectively:
\begin{align}\label{iis}
I_1(A) &:= S(A) \ge 0 \nonumber \\
I_2(A:B)&:= S(A)+S(B) - S(AB) \ge 0 \nonumber \\
I_3(A:B:C)&:=S(A)+S(B)+S(C) - S(AB)-S(BC)-S(AC)  +S(ABC) \le0 \,.
\end{align}
One could ask whether it is possible to define an $n$-fold entanglement entropy $I_n$ obeying a similar inequality.

The obvious generalization of \rref{iis} to $n$ regions includes a sum over all possible combinations of regions with alternating signs
\be\label{in}
I_n(A:B:C:\cdots) = \sum_{\sigma} (-1)^{|\sigma|} S(\sigma)\,.
\ee
Here the sum is over all possible subsets $\sigma$ of $\{A, B,C,\ldots \}$. (In the classical information-theory context, $I_n$ is called the $n$-dimensional interaction information \cite{McGill54}.) We note that this particular linear combination of entanglement entropies has the property that the short-distance divergences maximally cancel. For example, area law divergences for separated regions cancel in $I_n$ for $n>1$, because each region appears the same number of times in even and odd combinations. When two regions $A,B$ are adjacent, the area-law divergence for the shared part of their boundaries also cancels in $I_n$ for $n>2$, since $A$ occurs without $B$ (and $B$ without $A$) the same number of times in even and odd combinations. This was the motivation for the use of $I_3$ to compute the topological entanglement entropy in \cite{KitaevPreskill2006}.  With the definition \rref{in} this property generalizes in the natural way. When three regions $A,B,C$ meet along a codimension two three-fold corner, the divergences due to the corner in $A$ cancel in $I_n$ for $n>3$, since $A$ appears without either $B$ or $C$ the same number of times in even and odd combinations; similarly for the corners of $B$, $C$, $AB$, $AC$, and $BC$. More generally, the codimension-$m$ edges, which occur generically in $D>m$ dimensions, carry divergences that cancel in $I_n$ for $n>m+1$.

Unfortunately, however, the quantity \rref{in} does not obey any obvious inequality. One can find explicit examples of regions on the boundary of AdS${}_3$ where the $I_n$ take either sign for $n= 4,5$. We leave the question of other possible inequalities as a challenge for the future.

\subsection{Finite-$N$ corrections}

Monogamy is a property of holographic theories in the large-$N$ limit.  We suspect that this property will not in general persist at finite $N$ once bulk quantum effects become important.    To see this, let us consider a set of regions in a two-dimensional conformal field theory where the large-$N$ (classical) part of $I_3$ vanishes.  The first quantum correction to $I_3$ was computed in Section 2 and shown to be positive if the theory has a non-trivial operator with scaling dimension $\hat d<1/2$.  Many CFTs with low-dimension operators exist and may even have bulk duals.  For example, the ${\cal W}_N$ minimal models have operators with dimension $1/N$ and are conjectured to be dual to higher spin versions of three dimensional gravity coupled to matter \cite{Gaberdiel:2010pz}.  More generally, one can try to interpret the $N$-fold symmetric product of any CFT (and in particular one with an operator of dimension less than $1/2$) as a bulk theory of gravity. The bulk dual should be non-local, as can be seen by noting that the low-energy density of states has Hagedorn behaviour in the large-$N$ limit \cite{Keller:2011xi}.   Neither of these examples involves bulk theories described by classical Einstein gravity, where the RT formula applies.  A generalization of the RT formula to these cases  is outside the scope of this paper.  But these examples do suggest that it is possible for quantum effects in the bulk to spoil monogamy.  It would be interesting to investigate this question further.

\acknowledgments
We thank K. Br\'adler, H. Casini, K. Dasgupta, A. Kitaev, J. Oppenheim, M. van Raamsdonk, S. Shenker and E. Silverstein for useful discussions.  We also thank the following institutions where much of this work was carried out for their hospitality: Harvard University, the Institute for Physics and Mathematics of the Universe of Tokyo University, McGill University, Stanford University, and the Aspen Center for Physics. The work of PH and AM is supported by the Natural Sciences and Engineering Research Council of Canada. PH also acknowledges support from CIFAR, INTRIQ, the Perimeter Institute and ONR through grant No.\ N000140811249. The work of MH is supported in part by DOE Award No.\ DE-FG02-92ER40706.

\bibliography{superextensivity}

\providecommand{\href}[2]{#2}\begingroup\raggedright\begin{thebibliography}{10}

\bibitem{Ryu:2006bv}
S.~Ryu and T.~Takayanagi, {\it Holographic derivation of entanglement entropy
  from {AdS/CFT}},  {\em Physical Review Letters} {\bf 96} (2006) 181602,
  [\href{http://xxx.lanl.gov/abs/arXiv:hep-th/0603001}{{\tt
  arXiv:hep-th/0603001}}].

\bibitem{Ryu:2006ef}
S.~Ryu and T.~Takayanagi, {\it Aspects of holographic entanglement entropy},
  {\em Journal of High Energy Physics} {\bf 0608} (2006) 045,
  [\href{http://xxx.lanl.gov/abs/arXiv:hep-th/0605073}{{\tt
  arXiv:hep-th/0605073}}].

\bibitem{Nishioka:2009un}
T.~Nishioka, S.~Ryu, and T.~Takayanagi, {\it Holographic entanglement entropy:
  An overview},  {\em Journal of Physics A} {\bf A42} (2009) 504008,
  [\href{http://xxx.lanl.gov/abs/arXiv:0905.0932}{{\tt arXiv:0905.0932}}].

\bibitem{Headrick:2010zt}
M.~Headrick, {\it Entanglement {R}enyi entropies in holographic theories},
  {\em Physical Review} {\bf D82} (2010) 126010,
  [\href{http://xxx.lanl.gov/abs/arXiv:1006.0047}{{\tt arXiv:1006.0047}}].

\bibitem{Groisman2005}
B.~{Groisman}, S.~{Popescu}, and A.~{Winter}, {\it Quantum, classical, and
  total amount of correlations in a quantum state},  {\em Phys. Rev. A} {\bf
  72} (2005), no.~3 032317--+,
  [\href{http://xxx.lanl.gov/abs/arXiv:quant-ph/0410091}{{\tt
  arXiv:quant-ph/0410091}}].

\bibitem{Headrick:2007km}
M.~Headrick and T.~Takayanagi, {\it A holographic proof of the strong
  subadditivity of entanglement entropy},  {\em Physical Review} {\bf D76}
  (2007) 106013, [\href{http://xxx.lanl.gov/abs/arXiv:0704.3719}{{\tt
  arXiv:0704.3719}}].

\bibitem{CasiniHuerta2009}
H.~{Casini} and M.~{Huerta}, {\it Remarks on the entanglement entropy for
  disconnected regions},  {\em Journal of High Energy Physics} {\bf 3} (2009)
  48--+, [\href{http://xxx.lanl.gov/abs/arXiv:0812.1773}{{\tt
  arXiv:0812.1773}}].

\bibitem{Yeung1991}
R.~W. Yeung, {\it A new outlook on {S}hannon's information measures},  {\em
  {IEEE} Transactions on Information Theory} {\bf 37} (1991), no.~3 466--474.

\bibitem{McGill54}
W.~J. McGill, {\it Multivariate information transmission},  {\em Psychometrika}
  {\bf 19} (1954), no.~2.

\bibitem{KitaevPreskill2006}
A.~{Kitaev} and J.~{Preskill}, {\it Topological entanglement entropy},  {\em
  Physical Review Letters} {\bf 96} (2006), no.~11 110404--+,
  [\href{http://xxx.lanl.gov/abs/arXiv:hep-th/0510092}{{\tt
  arXiv:hep-th/0510092}}].

\bibitem{LindenWinter2005}
N.~Linden and A.~Winter, {\it A new inequality for the von {N}eumann entropy},
  {\em Communications in Mathematical Physics} {\bf 259} (2005) 129--138,
  [\href{http://xxx.lanl.gov/abs/arXiv:quant-ph/0406162}{{\tt
  arXiv:quant-ph/0406162}}].

\bibitem{Cadney2011}
J.~Cadney, N.~Linden, and A.~Winter, {\it Infinitely many constrained
  inequalities for the von neumann entropy},  {\em IEEE Trans. Inf. Theory}
  {\bf 58} (2012) 3657, [\href{http://xxx.lanl.gov/abs/arXiv:1107.0624}{{\tt
  arXiv:1107.0624}}].

\bibitem{Mark11}
V.~Balasubramanian, M.~B. McDermott, and M.~van Raamsdonk, {\it Momentum-space
  entanglement and renormalization in quantum field theory},
  \href{http://xxx.lanl.gov/abs/arXiv:1108.3568}{{\tt arXiv:1108.3568}}.

\bibitem{Casini:2004bw}
H.~Casini and M.~Huerta, {\it A finite entanglement entropy and the c-theorem},
   {\em Physics Letters} {\bf B600} (2004) 142--150,
  [\href{http://xxx.lanl.gov/abs/arXiv:hep-th/0405111}{{\tt
  arXiv:hep-th/0405111}}].

\bibitem{CasiniHuerta2005}
H.~{Casini}, C.~D. {Fosco}, and M.~{Huerta}, {\it Entanglement and alpha
  entropies for a massive dirac field in two dimensions},  {\em Journal of
  Statistical Mechanics: Theory and Experiment} {\bf 7} (2005) 7--+,
  [\href{http://xxx.lanl.gov/abs/arXiv:cond-mat/0505563}{{\tt
  arXiv:cond-mat/0505563}}].

\bibitem{Calabrese:2009ez}
P.~Calabrese, J.~Cardy, and E.~Tonni, {\it Entanglement entropy of two disjoint
  intervals in conformal field theory},  {\em Journal of Statistical Mechanics}
  {\bf 0911} (2009) P11001,
  [\href{http://xxx.lanl.gov/abs/arXiv:0905.2069}{{\tt arXiv:0905.2069}}].

\bibitem{Calabrese:2010he}
P.~Calabrese, J.~Cardy, and E.~Tonni, {\it Entanglement entropy of two disjoint
  intervals in conformal field theory {II}},  {\em Journal of Statistical
  Mechanics} {\bf 1101} (2011) P01021,
  [\href{http://xxx.lanl.gov/abs/arXiv:1011.5482}{{\tt arXiv:1011.5482}}].

\bibitem{HiaiOT81}
F.~Hiai, M.~Ohya, and M.~Tsukada, {\it {KMS} condition and relative entropy in
  von {N}eumann algebras},  {\em Pacific Journal of Mathematics} {\bf 96}
  (1981) 99--109.

\bibitem{Wolf2007}
M.~M. {Wolf}, F.~{Verstraete}, M.~B. {Hastings}, and J.~I. {Cirac}, {\it Area
  laws in quantum systems: Mutual information and correlations},  {\em Physical
  Review Letters} {\bf 100} (2008), no.~7 070502--+,
  [\href{http://xxx.lanl.gov/abs/arXiv:0704.3906}{{\tt arXiv:0704.3906}}].

\bibitem{LevinWen}
M.~{Levin} and X.-G. {Wen}, {\it Detecting topological order in a ground state
  wave function},  {\em Physical Review Letters} {\bf 96} (2006), no.~11
  110405--+, [\href{http://xxx.lanl.gov/abs/arXiv:cond-mat/0510613}{{\tt
  arXiv:cond-mat/0510613}}].

\bibitem{Pakman:2008ui}
A.~Pakman and A.~Parnachev, {\it Topological entanglement entropy and
  holography},  {\em Journal of High Energy Physics} {\bf 0807} (2008) 097,
  [\href{http://xxx.lanl.gov/abs/arXiv:0805.1891}{{\tt arXiv:0805.1891}}].

\bibitem{Fursaev:2006ih}
D.~V. Fursaev, {\it Proof of the holographic formula for entanglement entropy},
   {\em Journal of High Energy Physics} {\bf 0609} (2006) 018,
  [\href{http://xxx.lanl.gov/abs/arXiv:hep-th/0606184}{{\tt
  arXiv:hep-th/0606184}}].

\bibitem{deBoer:2011wk}
J.~de~Boer, M.~Kulaxizi, and A.~Parnachev, {\it Holographic entanglement
  entropy in {L}ovelock gravities},
  \href{http://xxx.lanl.gov/abs/arXiv:1101.5781}{{\tt arXiv:1101.5781}}.

\bibitem{Hung:2011xb}
L.-Y. Hung, R.~C. Myers, and M.~Smolkin, {\it On holographic entanglement
  entropy and higher curvature gravity},  {\em Journal of High Energy Physics}
  {\bf 04} (2011) 025, [\href{http://xxx.lanl.gov/abs/arXiv:1101.5813}{{\tt
  arXiv:1101.5813}}].

\bibitem{Yaffe:1981vf}
L.~G. Yaffe, {\it Large ${N}$ limits as classical mechanics},  {\em Reviews of
  Modern Physics} {\bf 54} (1982) 407.

\bibitem{Sachdev}
M.~A. {Metlitski}, C.~A. {Fuertes}, and S.~{Sachdev}, {\it {Entanglement
  entropy in the O(N) model}},  {\em Physical Review B} {\bf 80} (2009), no.~11
  115122--+, [\href{http://xxx.lanl.gov/abs/arXiv:0904.4477}{{\tt
  arXiv:0904.4477}}].

\bibitem{Petz86}
D.~Petz, {\it Sufficient subalgebras and the relative entropy of states of a
  von {N}eumann algebra},  {\em Communications of Mathematical Physics} {\bf
  15} (1986), no.~1 123--131.

\bibitem{HaydenJPW04}
P.~{Hayden}, R.~{Jozsa}, D.~{Petz}, and A.~{Winter}, {\it Structure of states
  which satisfy strong subadditivity of quantum entropy with equality},  {\em
  Communications in Mathematical Physics} {\bf 246} (2004) 359--374,
  [\href{http://xxx.lanl.gov/abs/arXiv:quant-ph/0304007}{{\tt
  arXiv:quant-ph/0304007}}].

\bibitem{Shirokov11}
M.~E. Shirokov Private communication, 2011.

\bibitem{SummersWerner87}
S.~J. {Summers} and R.~{Werner}, {\it {Bell's inequalities and quantum field
  theory. I. General setting}},  {\em Journal of Mathematical Physics} {\bf 28}
  (1987) 2440--2447.

\bibitem{HHH2000}
M.~{Horodecki}, P.~{Horodecki}, and R.~{Horodecki}, {\it Limits for
  entanglement measures},  {\em Physical Review Letters} {\bf 84} (2000)
  2014--2017, [\href{http://xxx.lanl.gov/abs/arXiv:quant-ph/9908065}{{\tt
  arXiv:quant-ph/9908065}}].

\bibitem{BennettDSW96}
C.~H. {Bennett}, D.~P. {Divincenzo}, J.~A. {Smolin}, and W.~K. {Wootters}, {\it
  Mixed-state entanglement and quantum error correction},  {\em Physical Review
  A} {\bf 54} (1996) 3824--3851,
  [\href{http://xxx.lanl.gov/abs/arXiv:quant-ph/9604024}{{\tt
  arXiv:quant-ph/9604024}}].

\bibitem{Narnhofer02}
H.~{Narnhofer}, {\it Entanglement, split and nuclearity in quantum field
  theory},  {\em Reports on Mathematical Physics} {\bf 50} (2002) 111--123.

\bibitem{KoashiWinter03}
M.~{Koashi} and A.~{Winter}, {\it Monogamy of quantum entanglement and other
  correlations},  {\em Physical Review A} {\bf 69} (2004) 022309--+,
  [\href{http://xxx.lanl.gov/abs/arXiv:quant-ph/0310037}{{\tt
  arXiv:quant-ph/0310037}}].

\bibitem{DevetakWinter05}
I.~{Devetak} and A.~{Winter}, {\it Distillation of secret key and entanglement
  from quantum states},  {\em Proceedings of the Royal Society of London Series
  A} {\bf 461} (2005) 207--235,
  [\href{http://xxx.lanl.gov/abs/arXiv:quant-ph/0306078}{{\tt
  arXiv:quant-ph/0306078}}].

\bibitem{ChristandlW04}
M.~{Christandl} and A.~{Winter}, {\it {``Squashed entanglement'': An additive
  entanglement measure}},  {\em Journal of Mathematical Physics} {\bf 45}
  (2004) 829--840, [\href{http://xxx.lanl.gov/abs/arXiv:quant-ph/0308088}{{\tt
  arXiv:quant-ph/0308088}}].

\bibitem{BrandaoCY10}
F.~G.~S.~L. {Brandao}, M.~{Christandl}, and J.~{Yard}, {\it Faithful squashed
  entanglement},  {\em Communications in Mathematical Physics} {\bf 306} (2011)
  805, [\href{http://xxx.lanl.gov/abs/arXiv:1010.1750}{{\tt arXiv:1010.1750}}].

\bibitem{Gaberdiel:2010pz}
M.~R. Gaberdiel and R.~Gopakumar, {\it {An AdS3 Dual for Minimal Model CFTs}},
  {\em Phys.Rev.} {\bf D83} (2011) 066007,
  [\href{http://xxx.lanl.gov/abs/1011.2986}{{\tt 1011.2986}}].

\bibitem{Keller:2011xi}
C.~A. Keller, {\it {Phase transitions in symmetric orbifold CFTs and
  universality}},  {\em JHEP} {\bf 1103} (2011) 114,
  [\href{http://xxx.lanl.gov/abs/1101.4937}{{\tt 1101.4937}}].

\end{thebibliography}\endgroup
\bibliographystyle{JHEP}
 
\end{document}